## *"I Wanted to Predict Elections with Twitter and all I got was this Lousy Paper"*

# A Balanced Survey on Election Prediction using Twitter Data

Daniel Gayo-Avello

`dani@uniovi.es`

`@PFCdgayo`

Department of Computer Science - University of Oviedo (Spain)

May 1, 2012

### Abstract

*Predicting X from Twitter* is a popular fad within the Twitter research subculture. It seems both appealing and relatively easy. Among such kind of studies, electoral prediction is maybe the most attractive, and at this moment there is a growing body of literature on such a topic.

This is not only an interesting research problem but, above all, it is extremely difficult. However, most of the authors seem to be more interested in claiming positive results than in providing sound and reproducible methods.

It is also especially worrisome that many recent papers seem to only acknowledge those studies supporting the idea of Twitter predicting elections, instead of conducting a balanced literature review showing both sides of the matter.

After reading many of such papers I have decided to write such a survey myself. Hence, in this paper, every study relevant to the matter of electoral prediction using social media is commented.

From this review it can be concluded that the predictive power of Twitter regarding elections has been greatly exaggerated, and that hard research problems still lie ahead.

# Introduction

In the last two years a number of papers have suggested that Twitter data has an impressive predictive power. Apparently, everything from the stock market [3], to movie box performance [1], through pandemics [12] are amenable to such forecasting.

Among these topics there is one which is very keen to me: Elections. My position on this is crystal clear and it can be summarized as:

*"No, you cannot predict elections with Twitter".*

I have discussed on print the many biases in the data and the flaws in current research on this topic [8] and I have joined forces with Takis Metaxas and Eni Mustafaraj to put to test the reproducibility of prior research on the area [14].

Needless to say, I'm far from being the only one skeptical about the feasibility of the methods proposed up to date. For instance, Jungherr *et al.* have criticized several flaws [10] in one of the most highly cited studies claiming that prediction is not only possible but even easy [23].

Those references are below for your reading pleasure but I will assume you are a busy person and, therefore, I will summarize the main problems with current "state of the art" election prediction using Twitter data:

## Flaws in Current Research regarding Electoral Predictions using Twitter Data

1. *It's not prediction at all!* I have not found a single paper predicting a future result. All of them claim that a prediction could have been made; i.e. they are *post-hoc* analysis and, needless to say, negative results are rare to find.

2. Chance is not a valid baseline because incumbency tends to play a major role in most of the elections.

3. There is not a commonly accepted way of "counting votes" in Twitter: current research has used raw volume of tweets, unique users, and many flavors of sentiment analysis.

4. *There is not a commonly accepted way of interpreting reality!* There are papers comparing the predicted results with polls, with popular vote, with the percentage of representatives each party achieves, etc.

5. Sentiment analysis is applied as a black-box and with naïveté. Indeed, most of the time sentiment-based classifiers are only slightly better than random classifiers.

6. All the tweets are assumed to be trustworthy. That is, the presence of rumors, propaganda and misleading information is ignored.

7. Demographics are neglected. Not every age, gender, social, or racial group is equally represented in Twitter. For instance, younger people from urban areas were overrepresented in the dataset I collected for the US 2008 Presidential elections and those people were clearly pro-Obama.

8. Self-selection bias is simply ignored. People tweet on a voluntary basis and, therefore, data is produced only by those politically active.

A number of tips can be drawn from this list:

# Recommendations for Future Research regarding Electoral Predictions using Twitter Data

1. There are elections virtually all the time, thus, *if you are claiming you have a prediction method you should predict an election in the future!*

2. Check the degree of influence incumbency plays in the elections you are trying to predict. Your baseline should not be chance but predicting the incumbent will win. Apply that baseline to prior elections; if your method's performance is not substantially better than the baseline then, sorry, you have a convoluted Rube Goldberg version of the baseline.

3. Clearly define which is a "vote" and provide sound and compelling arguments supporting your definition. For instance, why are you using all of the users even if they have just a few tweets on the topic? Or, why are you dropping users because they have few tweets on the topic? I know, it is hard and probably not even fair but we need to know the way votes are to be counted...

4. Clearly define the golden truth you are using. Again, sound and compelling arguments are needed but, in my opinion, you should use the "real thing" (i.e. avoid polls).

5. Sentiment analysis is a core task. We should not rely on simplistic assumptions and instead devote resources to the special case of sentiment analysis in politics before trying to predict elections.

6. Credibility should be a major concern. There is an incipient body of work in this regard (e.g. [5, 16]) so you should, at least, apply the available methods to justify the data you are using have been checked for credibility, and that disinformation, puppets, and spammers have been removed.

7. You should adjust your prediction according to (i) the participation of the different groups in the prior election you are trying to predict and (ii) the belonging of users to each of those groups. The second point is by far the hardest but you should try your best to obtain demographic data and political preference for the users in your dataset (cf. [6, 7, 15]).

8. The silent majority is a huge problem. Very little has been studied in this regard (see [17]) and this should be another central part of future research.

## Core Lines of Future Research

As I see it, the major lines of work regarding data mining of political tweets with forecasting purposes would be:

- Accurate sentiment analysis of political tweets. Please note that humor and sarcasm detection would play a major role here.

- Automatic detection of propaganda and disinformation.

- Automatic detection of sock puppets.

- Credibility checking.

- Basic research on Twitter demographics and automatic profiling of users with regards to demographic attributes.

- Basic research on user participation and self-selection bias.

# Relevant Prior Art

## Annotated Bibliography

This section provides an annotated bibliography on the topic of electoral prediction from Twitter data. The order is chronological, there are cross-references between papers, and some of them are just provided for the sake of comprehensiveness.

In addition to that, seminal papers not entirely related to the topic are included plus papers on related areas such as credibility, rumors, and demographics of Twitter users.

**Modeling Public Mood and Emotion: Twitter Sentiment and Socio-Economic Phenomena**  Bollen, J., Pepe, A., and Mao, H. 2009. Arxiv paper: arXiv:0911.1583v1. A peer-reviewed (albeit shorter) version was published as Bollen, J., Mao, H., and Pepe. A. 2011. "Modeling Public Mood and Emotion: Twitter Sentiment and Socio-Economic Phenomena," in *Proceedings of the Fifth International AAAI Conference on Weblogs and Social Media.*

To the best of my knowledge this is the first paper about the application of mood (not sentiment) analysis to Twitter data. In this piece of research POMS (Profile of Mood States) is used to distill from tweets time series corresponding to the evolution of 6 different emotional attributes (namely, *tension, depression, anger, vigor, fatigue,* and *confusion*).

POMS is a psychometric instrument which provides a list of adjectives for which the patient has to indicate his or her level of agreement. Each adjective is related to a mood state and, therefore, that list can be exploited as the basis for a naïve mood analyzer of textual data.

Bollen *et al.* applied POMS in that way to found that socioeconomic turmoil caused significant (although delayed) fluctuations of the mood levels.

It must be noted that although Bollen *et al.* argued that Twitter data could then be used for predictive purposes this paper does not describe any predictive method. Besides, although the US 2008 Presidential campaign and Obama Election are used as an scenario, no conclusions are inferred regarding the predictability of elections.

**From Tweets to Polls: Linking Text Sentiment to Public Opinion Time Series** O'Connor, B., Balasubramanyan, R., Routledge, B.R., and Smith, N.A. 2010. In *Proceedings of the Fourth International AAAI Conference on Weblogs and Social Media*.

This is one of the earliest papers discussing the feasibility of using Twitter data as a substitute for traditional polls.

O'Connor *et al.* employed the subjectivity lexicon from Opinion Finder to determine both a positive and a negative score for each tweet in their dataset (in their approach tweets can be both positive and negative at the same time). Then, raw numbers of positive and negative tweets regarding a given topic are used to compute a sentiment score (the ratio between the number of positive and negative tweets). It must be noted that O'Connor *et al.* clearly stated that by simple manual inspection they found many examples of incorrectly detected sentiment.

Using this method, sentiment time series were prepared for a number of topics (namely, consumer confidence, presidential approval and US 2008 Presidential elections). According to O'Connor *et al.* both consumer confidence and presidential approval polls exhibited correlation with Twitter sentiment data computed with their approach. However, no correlation was found between electoral polls and Twitter sentiment data.

**Predicting Elections with Twitter: What 140 Characters Reveal about Political Sentiment** Tumasjan, A., Sprenger, T.O., Sandner, P.G., and Welpe, I.M. 2010. In *Proceedings of the Fourth International AAAI Conference on Weblogs and Social Media*.

In all probability this is the paper which started all the fuzz regarding predicting elections using Twitter.

The paper has two clearly distinct parts: In the first one LIWC (Linguistic Inquiry and Word Count) is used to perform a superficial analysis of the tweets related to different parties running for the German 2009 Federal election. It is the second part, however, which has made this paper a highly cited one. There, Tumasjan *et al.* state that the mere count of tweets mentioning a party or candidate accurately reflected the election results. Moreover, they claim that the MAE (Mean Absolute Error) of the "prediction" based on Twitter data was rather close to that of actual polls.

It must be noted that this paper was responded by Jungherr *et al.* [10] which was in turn responded by Tumasjan *et al.* [24]; both papers are discussed below.

**From obscurity to prominence in minutes: Political speech and real-time search** Mustafaraj, E., and Metaxas, P. 2010. In *Proceedings of Web Science: Extending the Frontiers of Society On-Line.*

This paper introduces the concept of *Twitter-bomb*: the use of fake accounts in Twitter to spread disinformation by "bombing" targeted users who, in turn, would retweet the message, thus, achieving viral diffusion.

The paper describes the way in which a smear campaign was orchestrated in Twitter by a Republican group against Democrat candidate Martha Coakley and how it was detected and aborted. These ideas would inspire the Truthy project [20].

**Understanding the Demographics of Twitter Users** Mislove, A., Lehmann, S., Ahn, Y.Y., Onnela, J.P., and Rosenquist, J.N. 2011. In *Proceedings of the Fifth International AAAI Conference on Weblogs and Social Media.*

This paper describes how a sample of Twitter users in the US is analyzed along three different axes, namely, geography, gender and race/ethnicity.

The methods applied are simple but quite compelling. All of the data was inferred from the user profiles: geographical information was obtained from the self-reported location, while gender and race/ethnicity were inferred from the user name. Gender was determined by means of the first name and statistical data from the US Social Security Administration. Race/ethnicity was derived from the last name and data from the US 2000 Census.

Needless to say, such methods are prone to error but it is probably rather tolerable and the conclusions drawn from the study seem very sensible: highly populated counties are overrepresented in Twitter, users are predominantly male, and Twitter is a non-random sample with regards to race/ethnicity.

Mislove *et al.* also concluded that *post-hoc* corrections based on the detected over- and under-representation of different groups could be applied to improve predictions based on Twitter data. This is consistent with some of the findings of Gayo-Avello [8] regarding the correction of voting predictions on the basis of users age.

**Detecting and Tracking Political Abuse in Social Media** Ratkiewicz, J., Conover, M.D., Meiss, M., Gonçalves, B., Flammini, A., and Menczer, F. 2011. In *Proceedings of the Fifth International AAAI Conference on Weblogs and Social Media.*

This papers describes the Truthy project which was inspired by the *Twitter-bomb* concept coined by Mustafaraj and Metaxas [18].

Truthy is a system to detect astroturf political campaigns either to simulate widespread support for a candidate or to spread disinformation. The paper describes in reasonable detail the system, providing a number of real case examples and performance analysis.

**How (Not) To Predict Elections** Metaxas, P.T., Mustafaraj, E., and Gayo-Avello, D. 2011. In *Proceedings of PASSAT/SocialCom.*

One of the few papers casting doubts on the presumed predictive powers of Twitter data regarding elections and suggesting the use of incumbency as a baseline for predictions.

By analyzing results from a number of different elections Metaxas *et al.* concluded that Twitter data is only slightly better than chance when predicting elections.

In addition to that they describe three necessary standards for any method claiming predictive power on the basis of Twitter data: (1) it should be a clearly defined algorithm, (2) it should take into account the demographic differences between Twitter and the actual population, and (3) it should be "explainable", i.e. black-box approaches should be avoided.

**The Party Is Over Here: Structure and Content in the 2010 Election**
Livne, A., Simmons, M.P., Adar, E., Adamic, L.A. 2011. In *Proceedings of the Fifth International AAAI Conference on Weblogs and Social Media*.

This paper describes a method to predict elections which relies not only on Twitter data but also in additional information such as the party a candidate belongs to, or incumbency.

Livne *et al.* claim an 88% precision when incorporating Twitter data versus a 81% with such data omitted. The improvement is not substantial although certainly noticeable. It must be noted, however, that elections are modeled as processes with a binary outcome thus missing very important information, specially regarding tight elections or scenarios where coalitions are a possibility.

**Why the Pirate Party Won the German Election of 2009 or The Trouble With Predictions: A Response to Tumasjan, A., Sprenger, T. O., Sander, P. G., & Welpe, I. M. "Predicting Elections With Twitter: What 140 Characters Reveal About Political Sentiment"**
Jungherr, A., Jürgens, P., and Schoen, H. 2011. In *Social Science Computer Review*.

This paper is a response to that of Tumasjan *et al.* [23] previously mentioned. As it was explained, that paper claimed that the mere number of tweets is a good predictor for elections and that the distribution of tweets closely follow the distribution of votes for the different parties.

Jungherr *et al.* pointed out that the method by Tumasjan *et al.* was based on arbitrary choices (e.g. not taking into account all the parties running for the elections but just those represented in congress) and, moreover, its results varied depending on the time window used to compute them.

On the basis of those findings Jungherr *et al.* refuted the claim by Tumasjan *et al.* It must be noted that this was later responded by Tumasjan *et al.* in the following paper.

**Where There is a Sea There are Pirates: Response to Jungherr, Jurgens, and Schoen**   Tumasjan, A., Sprenger, T.O., Sandner, P.G., and Welpe, I.M. 2011. In *Social Science Computer Review*.

This paper is a response to the response by Jungherr *et al.* to the highly cited paper by Tumasjan *et al.* [23].

In the original paper Tumasjan *et al.* claimed impressive predictive powers for tweet counts; later, Jungherr *et al.* raised serious doubts on such a claim and in this paper Tumasjan *et al.* tried to dispel them.

Unfortunately, the arguments they provide in this paper are not compelling enough and, besides, they try to tone down their previous conclusions: saying that Twitter data is not to replace polls but to complement them; or stating that the prediction method was not the main finding of their original paper[1].

**On Using Twitter to Monitor Political Sentiment and Predict Election Results** Bermingham, A., and Smeaton, A.F. 2011. In *Proceedings of the Workshop on Sentiment Analysis where AI meets Psychology (SAAIP)*.

This paper discusses to a certain extent different approaches to incorporate sentiment analysis to a predictive method. Bermingham and Smeaton put their method to test with the 2011 Irish General Election finding that it is not competitive when compared with traditional polls.

It must be noted, however, that the method described in this paper is trained using polling data for the elections it aims to predict; therefore, it is debatable the point of a predictive method underperforming the results obtained from the training data.

**Don't turn social media into another 'Literary Digest' Poll** Gayo-Avello, D. 2011. In *Communications of the ACM*.

This papers discusses in detail the reasons for a failure predicting the outcome of the 2008 US Presidential Election. The author describes how different simple methods failed by predicting a Obama win... In every state, even Texas! The studied methods were analogous to those by Tumasjan *et al.* [23] or O'Connor *et al.* [19] and the author conducts a post-mortem on the different reasons for such a failure.

Hence, a number of problems are suggested: (1) The "file-drawer" effect, i.e. research with negative results refuting positive results are rarely published; (2) Twitter data is biased and it is not a representative sample; and (3) the sentiment analysis methods commonly used are naïve and not better than random classifiers.

**Vocal Minority versus Silent Majority: Discovering the Opinions of the Long Tail** Mustafaraj, E., Finn, S., Whitlock, C., and Metaxas, P.T. 2011. In *Proceedings of PASSAT/SocialCom*.

This paper provides compelling evidence on the existence of two extremely different behaviors in social media: on one hand there is a minority of users producing most of the content (*vocal minority*) and on the other there is a majority of users who hardly produce any content (*silent majority*).

---

[1]Even when the phrase "predicting elections with Twitter" prominently appears in the title of that paper.

With regards to politics these two groups are clearly separated and the vocal minority behaves as a resonance chamber spreading information aligned with their own opinions.

Because of such results Mustafaraj *et al.* suggest that extreme caution should be taken when building predictive models based on social media.

**Information Credibility on Twitter**  Castillo, C., Mendoza, M., and Poblete, B. 2011. In *Proceedings of WWW 2011*.

To the best of my knowledge this is the first paper describing an automatic method to separate credible from not credible tweets. It is somewhat related to the work by Mustafaraj and Metaxas [18] and Ratkiewicz *et al.* [20].

The paper describes in detail the features to extract from the tweets to then obtain a decision tree. According to Castillo *et al.* such classifier showed a performance comparable to other machine learning methods –albeit slightly better. The work by Morris *et al.* [16] (see below) is certainly related to this one.

**Predicting the 2011 Dutch Senate Election Results with Twitter**  Tjong, E., Sang, K., and Bos, J. 2012. In *Proceedings of SASN 2012, the EACL 2012 Workshop on Semantic Analysis in Social Networks*.

In this paper Twitter data regarding the 2011 Dutch Senate elections was analyzed. Tjong *et al.* concluded that tweet counting is not a good predictor; therefore, contradicting the conclusion of Tumasjan *et al.* [23] and providing indirect support to the conclusions by Jungherr *et al.* [10], Metaxas *et al.* [14], or Gayo-Avello [8]. They also found that applying sentiment analysis can improve performance, a result consistent again with those by Metaxas *et al.* or Gayo-Avello.

Nevertheless, the performance of their method is below that of traditional polls and, in addition to that, the method relies on polling data to correct for demographic differences in the data. In this regard, it shares the same flaw of the work by Bermingham and Smeaton [2].

**Tweets and Votes: A Study of the 2011 Singapore General Election**  Skoric, M., Poor, N., Achananuparp, P., Lim, E-P., and Jiang, J. 2012. In *Proceedings of the 45th Hawaii International Conference on System Sciences*.

This paper is in line with the previous one: it shows that there is a certain amount of correlation between Twitter chatter and votes but that it is not enough to make accurate predictions. The performance found by these authors (using MAE as measure of performance) is much worse than the one reported by Tumasjan *et al.* [23] and they also found that although Twitter data can provide a more or less reasonable glimpse on national results it fails when focusing on local levels.

In addition to the technical caveats for this kind of predictions, the authors discuss some additional external problems affecting the predictive power of such methods: namely, democratic maturity of the country, competitiveness of the election, and media freedom.

**Tweeting is Believing? Understanding Microblog Credibility Perceptions** Morris, M.R., Counts, S., Roseway, A., Hoff, A., and Schwarz, J. 2012. In *Proceedings of CSCW 2012.*

This paper is highly related to that by Castillo *et al.* [5]. Nevertheless, Morris *et al.* did not develop an automatic way to determine the credibility of a tweet but they conducted a survey to find the features that make users to perceive a tweet as credible. They found that content alone is not enough to assess truthfulness and that users rely on additional heuristics. Such heuristics can be manipulated by the authors of tweets and, therefore, they can affect credibility perceptions for the better and for the worse.

## Additional Bibliography

The following list of papers is not related to electoral prediction using social media but they can provide the reader a broader perspective for that topic:

- Asur, S., and Huberman, B.A. 2010. **Predicting the Future With Social Media**. In *Proceedings of the 2010 IEEE/WIC/ACM International Conference on Web Intelligence and Intelligent Agent Technology* - Volume 01.

- Conover, M.D., Ratkiewicz, J., Francisco, M., Gonçalves, B., Flammini, A., and Menczer, F. 2011. **Political Polarization on Twitter**. In *Proceedings of the Fifth International AAAI Conference on Weblogs and Social Media.*

- Golbeck, J., Hansen, D.L. 2011. **Computing Political Preference among Twitter Followers**. In *Proceedings of the 2011 annual conference on Human factors in computing systems.*

- Conover. M.D., Gonçalves, B., Ratkiewicz, J., Flammini, A., and Menczer, F. 2011. **Predicting the Political Alignment of Twitter Users**. In *Proceedings of 3rd IEEE Conference on Social Computing SocialCom.*

- Jürgens, P., Jungherr, A., and Schoen, H. 2011. **Small Worlds with a Difference: New Gatekeepers and the Filtering of Political Information on Twitter**. In *Proceedings of the 3rd ACM International Conference on Web Science.*

- Yu, S., and Kak, S. 2012. **A Survey of Prediction Using Social Media**. Arxiv paper: arXiv:1203.1647v1.

# Conclusion and Final Recommendations

From the previous literature review it can be concluded that:

1. *Not everybody is using Twitter.* Social media is not a representative and unbiased sample of the voting population. Some strata are underrepresented while others are over-represented in Twitter. Demographic bias should be acknowledged and predictions corrected on its basis.

2. *Not every twitterer is tweeting about politics.* A minority of users are responsible for most of the political chatter and, thus, their opinions will drive what can be predicted from social media. This self-selection bias is still an open problem.

3. *Just because it is on Twitter does not mean it is true.* A substantial amount of data is not trustworthy and, thus, it should be discarded. There is a growing body of work in this regard but it is not being widely applied when trying to predict elections.

4. *Naïveté is not bliss.* Simplistic sentiment analysis methods should be avoided one and for all. Political discourse is plagued with humor, double entendres, and sarcasm; this makes determining political preference of users hard and inferring voting intention even harder.

Therefore, if you are planning to conduct serious research in this topic please take into account all of that; try to follow some of the research lines I have outlined; and, above all, do not cherry-pick references to support your point because, remember, *you cannot (consistently) predict elections from Twitter!*